\documentclass[a4paper,twoside]{article}
\usepackage{pasa}
\usepackage{graphicx}
\usepackage{natbib}
\usepackage{color}

\newcommand{\vex}{\vspace{1ex}}
\newcommand{\vhx}{\vspace{0.5ex}}

\newcommand{\ntab}[2]{ \multicolumn{1}{#1}{#2} }

\newcommand{\nnntab}[2]{ \multicolumn{3}{#1}{#2} }

\newcommand{\dss}{\displaystyle}
\newcommand{\lp}{ \left(  }
\newcommand{\rp}{ \right) }

\newcommand{\Frac}[2]{\frac{\displaystyle\strut #1}{\displaystyle\strut #2} }

\newcommand{\getlength}[1]{\ifx#1\end \let\next=\relax
            \else\advance\count255 by1 \let\next=\getlength\fi \next}

\newcommand{\ifnularg}[1]{ \count255=0 \getlength#1\end \ifnum\count255=0 }
\newcommand{\ifm}{\makebox{}\ifmmode}
\long\def\ifundefined#1#2#3{\expandafter\ifx\csname
  #1\endcsname\relax#2\else#3\fi}
\newcommand{\beq}   { \begin{eqnarray} }
\newcommand{\eeq}[1]{ \ifnularg{#1} end{eanarray} \else
                      \label{#1}\end{eqnarray}    \fi }
\newcommand{\eeql}   { \end{eqnarray} }
\newcommand{\eeqn}   { \nonumber \end{eqnarray} }

\newcommand{\mnras}{MNRAS}
\newcommand{\pasp}{PASP}
\newcommand{\pasa}{PASA}
\newcommand{\jgr}{JGR}
\newcommand{\aj}{AJ}
\renewcommand{\aa}{A\&A}
\newcommand{\Tilde}{\char`\~}
\newcommand{\askap}{\mbox{\sc askap-\footnotesize{2\hspace{-0.01em}9}}}
\newcommand{\Askap}{{\sc askap}-{\footnotesize{2\hspace{-0.01em}9}}}
\newcommand{\wark}{\mbox{\sc wark\hspace{0.02em}{\footnotesize{1\hspace{-0.01em}2}}\hspace{-0.0em}m}}
\newcommand{\atca}{\mbox{\sc atcapn\hspace{0.03em}{\footnotesize{5}}}}
\newcommand{\hobart}{\mbox{\sc hobart\hspace{0.04em}{\footnotesize{2\hspace{-0.01em}6}}}}

\newcommand{\Degr}{\ifm{}^\circ\!\else${}^\circ\!$\fi}
\newcommand{\degr}{\ifm{}^\circ\else${}^\circ$\fi}
\newcommand{\PIMA}{$\cal P\hspace{-0.067em}I\hspace{-0.067em}M\hspace{-0.067em}A$ }
\newcommand{\Number}[1]{\ifnum#1<10\relax0\number#1\else\number#1\fi}
\newcommand{\isodate}{
\count151=\time
\divide\count151 by 60
\count151=\count151
\multiply\count151 by 60
\count152=\time
\advance\count152 by -\count151
\divide\count151 by 60
\count152=\count151
\multiply\count151 by 60
\count153=\time
\advance\count153 by -\count151
\Number{\year}.\Number{\month}.\Number{\day}--\Number{\count152}:\Number{\count153}
}

\normalfont

\definecolor{Dred}{rgb}{0.312,0.070,0.070}
\definecolor{Dblue}{rgb}{0.070,0.070,0.312}
\definecolor{Dgreen}{rgb}{0.070,0.312,0.070}
\definecolor{Db}{rgb}    {0.050,0.0,0.320}

\newcounter{note}
\let\oldmarginpar\marginpar
\renewcommand\marginpar[1]{\-\oldmarginpar[\raggedleft\footnotesize #1]%
{\raggedright\footnotesize #1}}
\newcommand{\Note}[1]{\Rdb{#1}{\addtocounter{note}{1}%
\marginpar{\small\underline{\Rdb{Corr \arabic{note}}}}}}
\newcommand{\note}[1]{\Rdb{#1}}
\renewcommand{\note}[1]{#1}
\renewcommand{\Note}[1]{#1}
\begin{document}

\title{First geodetic observations using new VLBI stations ASKAP-29 and WARK12M}
\ShortTitle{First geodetic observations at ASKAP-29 and WARK12M}
\RunningAuthors{L. Petrov et al.}
\PubYear{2011}
\DOI{10.1071/AS10048}
\JournalNumber{28(2)}
\PageNumber{107--116}
\setcounter{page}{107}

\NumberofInstitutions{6}
\InstitutionName{1}{ADNET Systems, Inc./NASA GSFC, Code 610.2, Greenbelt, 
                    MD 20771 USA}
\InstitutionName{2}{CSIRO Astronomy and Space Science, PO Box 76, 
                    Epping, NSW 1710, Australia}
\InstitutionName{3}{International Centre for Radio Astronomy Research, 
                    Curtin University, Bentley, Western Australia, 6102,
                    Australia}
\InstitutionName{4}{Institute for Radio Astronomy and Space Research, 
                    Auckland University of Technology, Private Bag 92006,
                    Auckland 1142, New Zealand}
\InstitutionName{5}{GNS Science, PO Box 30368, Lower Hutt 5040, New Zealand}
\InstitutionName{6}{Land Information New Zealand, 160 Lambton Quay, PO 
                    Box 5501, Wellington 6145, New Zealand}
\EmailAddress{Leonid.Petrov@lpetrov.net}
\EmailAddressLetter{G}

\AuthorList{
        Leonid Petrov\affil{A,G},
        Chris Phillips\affil{B},
        Tasso Tzioumis\affil{B},
        Bruce Stansby\affil{C},
        Cormac Reynolds\affil{C},
        Hayley E Bignall\affil{C},
        Sergei Gulyaev\affil{D},
        Tim Natusch\affil{D},
	Neville Palmer\affil{E},
	David Collett\affil{F},
        John E Reynolds\affil{B},	
        Shaun W Amy\affil{B},	
        Randall Wayth\affil{C},	
        Steven J Tingay\affil{C}
}

\ReceivedDate{December 20, 2010}
\AcceptedDate{February 22, 2011}
\PublishedDate{June 16, 2011}

\Abstract{
We report the results of a successful 7~hour 1.4~GHz VLBI experiment using
two new stations, ASKAP-29 located in Western Australia and
WARK12M located on the North Island of New Zealand. This was the first 
geodetic VLBI observing session with the participation of these new stations. 
We have determined the positions of ASKAP-29 and WARK12M. Random errors 
on position estimates are 150--200~mm for the vertical component and 40--50~mm 
for the horizontal component. Systematic errors caused by the unmodeled 
ionosphere path delay may reach 1.3~m for the vertical component.
}
\pasamaketitle
\thispagestyle{empty}

\section{Introduction} \label{s:intro}

As part of the Australian and New Zealand joint bid to host the
multi-billion dollar Square Kilometre Array (SKA), both countries are
investing heavily in advanced technologies for radio astronomy.  In
Australia this follows a strong tradition in radio astronomy and is
expressed in the construction of the Australian SKA Pathfinder (ASKAP)
on the western edge of the Australian continent  \citep{r:jonston08}.
Radio astronomy in New Zealand has links stretching back to the work 
of Elizabeth Alexander on solar radio emission \citep{r:ale46}.
John Bolton and Gordon Stanley used a cliff interferometer to obtain rising 
and setting records of various radio sources \citep{r:bol82}; 
measurements made near Sydney and Auckland allowed them to identify
mysterious ``radiostars'' with well known supernova remnants and galaxies 
\citep{r:bol49}. The first successful VLBI experiment between Australia 
and New Zealand was made in 2005 with a 6-m radio telescope near 
\Note{Auckland (Karaka) and} the Australia Telescope Compact Array (ATCA) 
\citep{r:gul05,r:tin06}. In 2008, Auckland University of Technology 
commissioned a new 12-m antenna at Warkworth near Auckland, --- the 
New Zealand's first research capable radio telescope \citep{r:gul09}.

A joint Australia and New Zealand collaborative project is focused on
developing both the ASKAP and Warkworth facilities as part of regional
and global Very Long Baseline Interferometry (VLBI) arrays, for the
purposes of both astronomy and geodesy.  Due to the dearth of land
mass in the Southern Hemisphere, both astronomical and geodetic VLBI
has suffered in the past, with the radio telescopes used for VLBI
restricted to the south east ``corner'' of the Australian continent, one
telescope in South Africa and occasional use of radio telescopes 
in Antarctica and South America.

\Note{The additional capability gained by adding radio telescopes located 
in Western Australia and New Zealand is substantial, increasing the 
angular resolution of the Australasian array by a factor of approximately
four}, admitting a range of astronomy science goals described in 
\citet{r:jonston08}.  


High precision astrometry is likely to be an important part of the
science case for the high angular resolution component of the SKA, in
particular for precise determination of the distances of radio pulsars
in the galaxy, to be used in various tests of fundamental physics
\citep{r:smits10}.  In order to achieve the
astrometric performance required, the SKA will need to undertake
astrometry utilising distributed clusters of small antennas operating
as phased arrays.  This is a departure from standard precision
astrometry and our work using ASKAP as part of a VLBI array for
astrometry will be an important testbed for the demonstration of these
techniques.

  The 12-m radio telescope \wark\footnote{\wark\  and 
\askap\ are identifiers for specific VLBI antennas: near Warkworth, 
New Zealand and at the Murchison Radio-Astronomy Observatory, Western
Australia, respectively.} is intended to 
be used as a part of the Australian Long Baseline Array (LBA), for spacecraft 
monitoring, and for VLBI observations in the framework of the 
International VLBI Service for Geodesy \& Astrometry 
(IVS)\footnote{{\tt http://ivscc.gsfc.nasa.gov/}} and the 
AuScope project\footnote{{\tt http://www.auscope.org.au/}}.

  \askap\ is an element of the ASKAP
array of 36 identical dishes using Phased Array Feed (PAF) technologies, 
expected to be fully operational by 2013.  ASKAP will undertake very wide 
field survey science in continuum and spectral line modes
and is also intended to be used as a part of regional and global VLBI 
networks for a variety of 
projects\footnote{{\tt http://www.atnf.csiro.au/SKA/}}.

  As a part of commissioning the new antennas, the positions of antenna reference
points should be determined. A reference point is defined as 
the point of the projection of the movable elevation axis onto the fixed 
azimuthal axis. For the analysis of VLBI source imaging experiments
\Note{made in a phase-referencing mode}, the projection of a baseline vector,
(i.e. vector between antenna reference points) to the tangential image plane 
should be known with errors not exceeding tens of centimetres, otherwise 
the image will be smeared \Note{\citet{r:cha2002}}. For astrometry 
applications, angular position accuracies of tens of $\mu$as are required,
making the requirements on the accuracy of station positions much more 
stringent: 5--10~mm. One way of estimating the position of the antenna 
reference points is through analysis of a combination of a ground survey 
of markers attached to the antennas from a local network around the station 
and GPS observations from the points at the local 
network \citep[see e.g.,][]{r:sarti04,r:sarti09}. Another way to estimate 
station positions is to use the VLBI technique itself to determine group 
delays and then derive reference point positions from these group 
delay measurements. The advantage of this approach is that it also 
provides useful diagnostics on the VLBI equipment.

  The first fringe test experiment between \Askap\ and \wark\ was made 
in April 2010. First fringes 
on baselines between \askap\ and {\sc mopra} were found 
on April 22, 2010 and on baselines to \wark\ on the following day. 
This success prompted three first science experiments using the full 
Long Baseline Array (LBA) network with the participation of the two 
new stations: 1)~imaging observations of {\tt PKS 1934$-$638} on 
29--30th April \citep{r:askap1}, 2)~a geodetic experiment on May 07, 2010, 
and 3)~imaging onbservations of {\tt Cen-A} on May 09, 2010 
(Tingay et al., 2010, in preparation). 

  We report here results from the geodetic experiment that was conducted at 
\askap\ and during the first geodetic observing session 
in May 07, 2010. The goal of this experiment was to determine the position of 
the antennas with decimeter accuracy and to collect diagnostic data. 
The characteristics of new antennas are presented in section \ref{s:ant}.
The experiment and its analysis are described in sections \ref{s:exp}
and \ref{s:anal}. Concluding remarks are made in section \ref{s:concl}.

\section{New Antennas}     
\label{s:ant}

\subsection{\sc askap-29}

\begin{table}
   \caption{\bf Specifications of an ASKAP antenna}
   \par\medskip\par
   \begin{tabular}{ll}
      \hline
      Antenna Type      & fully-steerable,      \\
                        &  prime focus          \\
      Reflector         & 12-metre paraboloid   \\
      Surface accuracy  & 0.5~mm rms or better  \\
      Mount             & 3 axes: polarization, \\
                        & elevation, azimuth    \\
      Height of elev axis & 7.51~m above \\
                          &     concrete foundation       \vex \\
      Azimuth axis range       &  $-180\degr$ to $+360\degr$   \\
      Azimuth axis speed       &  $3\degr$/s                   \\
      Azimuth axis accel.      &  $3\degr$/s$^2$          \vex \\
      Elevation axis range     &  $15\degr$ to $89\degr$       \\
      Elevation axis speed     &  $1\degr$/s                   \\
      Elevation axis accel.    &  $1\degr$/s$^2$          \vex \\
      Polarization axis range  &  $-180\degr$ to $+180\degr$   \\
      Polarization axis speed  &  $3\degr$/s                   \\
      Polarization axis accel. &  $3\degr$/s$^2$               \\
      \hline
  \end{tabular}
  \label{t:askap}
\end{table}

The Australian SKA Pathfinder (ASKAP) is an advanced technology radio 
telescope currently under construction at the Murchison Radio Observatory 
(MRO) in Western Australia. The ASKAP array comprises 36 fully-steerable dish 
antennas of 12 metre aperture fitted with phased array feed (PAF) 
receivers, giving an instantaneous field of view of approximately 30 square 
degrees with an instantaneous bandwidth of 300 MHz in the range 
700--1800~MHz.

The first antenna, \askap, was commissioned in March 2010 and 
was temporarily fitted with a conventional horn (single-pixel) L~band 
feed \Note{(1.4~GHz)} and uncooled receiver. Five additional antennas were 
completed in Q4 2010 with the balance to be commissioned throughout 2011.

The antenna is of a prime focus design with an unshaped paraboloidal 
reflector of f/D 0.5. The original specifications on surface accuracy
were 1~mm root mean square (rms). However, measurements showed that 
the surface accuracy of the \askap\ reflector and six other 
identical antennas manufactured by November 2010 is better than 0.5~mm 
rms over the full elevation range. There is some flexure in the position 
of the focus at a level of few mm over the full elevation range, 
but this can be compensated for if required. This level of surface accuracy 
implies that the antenna will be usable at frequencies as high as 
\Note{Ka~band (30~GHz) and possibly at Q-band (43~GHz).} A quadripod supports 
the prime focus platform with maximum  payload of about 200~kg. Additional 
bracing supports on each leg have been added to the original design to reduce 
flexure.

An unusual feature of the antenna is the three-axis mount. Above the 
familiar azimuth and elevation axes is the so-called polarization axis, 
allowing the entire reflector to be rotated about its optical axis. This 
design was specified to allow full parallactic angle tracking for high 
dynamic range in imaging observations, as it improves modeling of far-out 
sidelobes caused by the quadripod. 
An equatorial (``HA-Dec'') design naturally maintains a constant 
orientation of its focal plane with respect to the celestial sphere when 
tracking a fixed point in celestial coordinates. However this type of 
mount usually suffers restricted sky coverage due to limited rotation in 
hour angle.

Another feature of the antenna is the relatively large pedestal specified 
to accommodate the analogue and digital electronics for the PAF receiver 
with its 188 individual ports. The 54th Research Institute of China 
Electronics Technology Group Corporation (known as CETC54) are contracted 
to design, manufacture and install all thirty-six antennas at the MRO.

\begin{figure}[ht]
   \includegraphics[width=0.48\textwidth]{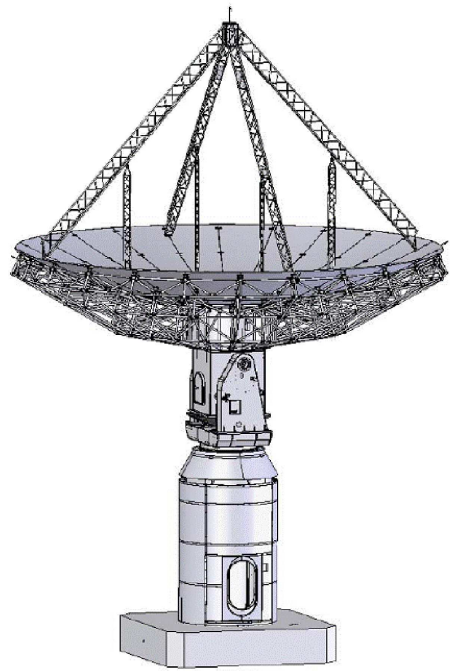}
   \par\vspace{-1ex}\par
   \caption{{\sc askap} antenna.}
           \label{f:askap}
   \par\vspace{-1ex}\par
\end{figure}

The baseband data at \askap\ were recorded directly to disk utilising  
a custom recording system developed at Curtin University. The recording 
systems were assembled from consumer off-the-shelf components. 
The recorders were interfaced with the CSIRO supplied analogue 
down-converters. The systems  recorded the data to disk at the 
native sampler/digitizer sample size of 16~bits.

  The 16~bit sample size gives an aggregate data rate when sampled at
the Nyquist-Shannon rate over 2~polarization channels of 4~Gbit/s. 
The recorder system comprised a Signatec PX14400 PCIe sampler/digitizer 
mounted in a server class computer. The systems were built in a 16 disk 
chassis populated with SATA hard drives. The recording system utilised 
a Linux software RAID, and the 16 disks where subdivided into 8~disk raid 
units. The recorded data were converted to the LBA format with 2~bit 
samples after the experiment using software employing automatic gain 
control in the conversion process in order to match the configuration 
of LBADR recorders used at other stations.

\subsection{WARK12M}

\begin{figure}[hbt]
   \includegraphics[width=0.48\textwidth]{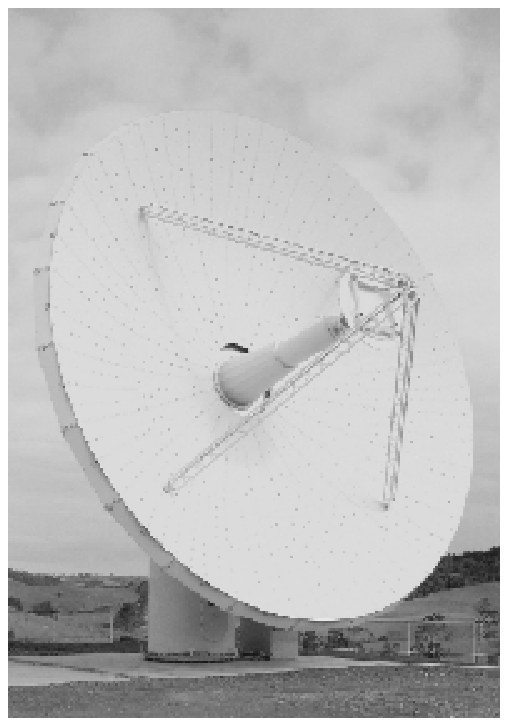}
   \caption{\wark\ antenna.}
           \label{f:wark12m}
\end{figure}

  The New Zealand 12-m radio telescope (see Figure~\ref{f:wark12m})
is located some 60 km north of the city of Auckland, near the township 
of Warkworth. It was manufactured by Patriot Antenna Systems 
(now Cobham Antenna Systems), USA.  The antenna specifications are provided 
in Table~\ref{t:wark}. The radio telescope was originally designed to 
operate at S \Note{(2.3~GHz)} and X~bands \Note{(8.4~GHz)} and it was 
supplied with an S/X dual-band dual-polarisation feed. 
Photogrammetry observations showed that the surface rms accuracy is 
0.35~mm. Therefore, the antenna is suitable for observations at frequencies
as high as 43~GHz. It is equipped with a digital base band converter 
(DBBC) developed by the Italian Institute of Radio Astronomy, Symmetricom 
Active Hydrogen Maser MHM-2010 (75001--114) and with the Mark5B+ data 
recorder developed at MIT Haystack Observatory. However these were  not 
installed at the time of this experiment, so data were recorded with the same 
equipment as installed at \askap.

\begin{table}
   \caption{\bf Specifications of an \wark\ antenna}
   \par\medskip\par
   \begin{tabular}{ll}
        \hline
        Antenna type & Fully-steerable, \\
                     & dual-shaped Cassegrain \\
        Manufacturer & Cobham/Patriot, USA \\
        Main dish Diam.      & 12.1 m  \\
        Secondary refl. Diam. & 1.8 m \\
        Focal length              & 4.538 m \\
        Surface accuracy          & 0.35 mm \\
        Pointing accuracy         & $18''$  \\
        Frequency range           & 1.4---43~GHz \\
        Mount                     & alt-azimuth \\
        Azimuth axis range        & $90\degr \pm 270 \degr$ \\
        Elevation axis range      & $4.5\degr$ to $88\degr$ \\
        Azimuth axis max speed    & $5\degr$/s \\
        Elevation axis max speed  & $1\degr$/s \\
        Main dish F/D ratio:      & 0.375 \\
        \hline
   \end{tabular}
   \label{t:wark}
\end{table}

  The antenna elevation axis is at a height of approximately 7.1~m 
above the ground level. The elevation axis is supported by a pedestal 
that is of steel construction. The pedestal is essentially a steel cylinder 
of  $\sim\!2.5$~m diameter. Apart from the pedestal all other 
components of the antenna (the reflector and feed support structure) are
constructed of aluminum. The support foundation for the antenna is 
a reinforced concrete pad that is 1.22~m thick by 
$6.7 \times 6.7$~meters square. The ground that the foundation is laid 
on consists of weathered sandstone/mudstone, i.e. is of sedimentary origin,
laid down in the Miocene period some 20~million years ago.  

  The radio telescope is directly connected to the regional advanced network 
KAREN (Kiwi Advanced Research and Education Network), which provides 
a fast connectivity between New Zealand's educational and research 
institutions \citep{r:karen}. At the time of the experiment, 
the Trans-Tasman (New Zealand--Australia) connectivity from the observatory
was 155~Mbps. By the end of 2010 this was upgraded to 1 Gbps.

  A preliminary survey has been conducted in collaboration with the New Zealand 
Crown Research Institute (CRI), GNS Science and Land Information New 
Zealand (LINZ) to determine an initial estimate of the reference point 
of the VLBI site \wark. This reference point is defined as the intersection 
of the azimuth and elevation axes of the telescope.  A real-time kinematic 
(RTK) GPS method was used to derive the position with respect to the GPS 
station WARK.

  WARK was established in November 2008 at the radio telescope site and 
is one of thirty nine PositioNZ network stations \citep{r:bli10} in New 
Zealand\footnote{http://www.linz.govt.nz/geodetic/positionz/}.  
All data received from the PositioNZ stations are compiled into 24~hour 
sessions and are processed to produce daily positions for each station 
in terms of ITRF2000\footnote{\parbox[t]{0.48\textwidth}{For more details see \\
{\tt http://geonet.org.nz/resources/gps/ \\
gps-processing-notes.html}}}.  The coordinates for WARK used in the following 
calculations were derived by averaging the daily coordinate solutions for 
February 19 through March 09, 2010.

  The RTK reference receiver was set up in an arbitrary location with 
clear sky view and was configured to record raw observations in addition 
to transmitting real-time corrections.  This station was later 
post-processed with respect to WARK and all RTK rover surveyed positions 
were subsequently adjusted relative to the updated reference position.

  Several points on the rim of the main reflector were identified and 
each point was measured several times with the RTK rover while the 
telescope was repositioned in elevation and azimuth between each 
measurement.  The rover GPS antenna was mounted on a 0.5 m survey pole 
and was hand held for each measurement.  Access to the rim of the reflector 
was achieved with a hydraulic cherry picker.

  The sequence of observation for determination of the horizontal axis was as follows: The telescope 
azimuth axis was held fixed (nominally $0^\circ$). A point near the highest 
edge of the reflector was identified and measured with the telescope in 
4 positions of elevation, from almost zenith ($\sim\!88^\circ$) to as high as 
the cherry picker could reach ($\sim\!38^\circ$). This was repeated 
a second time with a point identified on the edge of the reflector to one 
side of the telescope.  Five positions of elevation were measured at this 
point, from $10^\circ$ to $80^\circ$.

  The sequence of observation for determination of the vertical axis was as follows: The telescope 
elevation axis was held fixed (nominally $80^\circ$). Three points around 
the edge of the reflector were identified. The telescope azimuth axis was 
rotated into three positions such that each identified point could be 
measured consecutively from the surveyor's location in the cherry picker.
The cherry picker was then repositioned twice around the perimeter of the 
telescope and the measurements were repeated at each cherry picker location.  
This provided three measurements of each identified point with a fixed 
telescope zenith and varying azimuths.

  The resulting points from these measurements describe two vertical 
circles of rotation which define the movable elevation axis and three 
horizontal circles of rotation which define the fixed azimuthal axis.  
The coordinates for all subsequent calculations were retained as 
geocentric Cartesian coordinates to avoid any possibility of errors 
related to transformation of projection.

  The following method was used to determine the axes and their intersection 
point from these observations.   The equation of a circle from 
3~points (P.~Bourke, 1990, internal memo\footnote{Available at 
{\tt http://local.wasp.uwa.edu.au/\Tilde{pbourke/} \par geometry/circlefrom3}})
was used to calculate all possible combinations of three observed points 
which define a circle of rotation.  A simple mean was taken for all 
horizontal axis definitions and all vertical axis definitions.  The mid 
point of the closest point of approach of each axis with the other was 
used as the final estimate of the point of intersection.  The distance 
between closest point of approach on each axis was calculated to be 24~mm.  
Based on the variation of results for different combinations of survey 
points, we estimate that the accuracy of the determined intersection point 
is within 0.1~m.

In summary, the following coordinates of the intersection of the azimuth 
and elevation axes for the radio telescope \wark\ were derived in terms 
of ITRF2000 at the epoch of the survey (March 2010):
\beq
    \begin{array}{lcrll}
       X  & = &  -5115324.5 & \pm & 0.1 {\rm m} \\
       Y  & = &    477843.3 & \pm & 0.1 {\rm m} \\
       Z  & = &  -3767193.0 & \pm & 0.1 {\rm m} \\
    \end{array}
\eeqn

  It is intended that the radio telescope reference point coordinates 
will subsequently be re-determined to a higher accuracy with the use 
of a variety of terrestrial and GNSS survey techniques \citet{r:dawson04}
and a more rigorous least squares analysis of the observations.  Four geodetic 
survey monuments have been built within 15--20~m from the antenna pedestal for 
this purpose.

\section{Geodetic VLBI Experiment}  \label{s:exp}

\begin{figure}[ht]
   \includegraphics[width=0.48\textwidth]{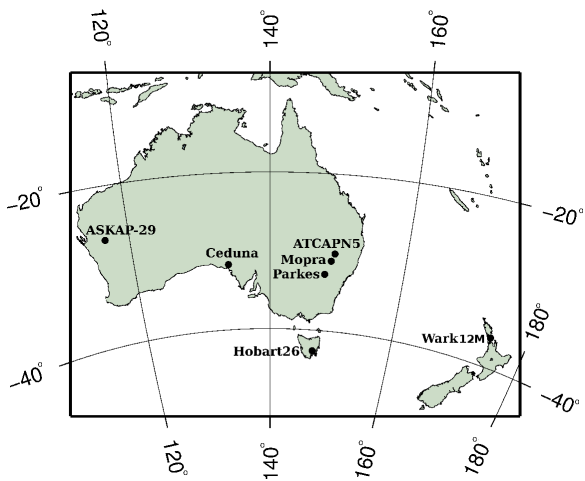}
   \caption{The LBA network used in vt14a experiment.}
   \label{f:lba_map}
\end{figure}

Seven stations participated in the 7~hour long VLBI experiment vt14a
on May 07, 2010, with the goal of determining station coordinates.
The stations are listed in Table~\ref{t:lba} and shown 
in Figure~\ref{f:lba_map}. Usually geodetic experiments are made in two 
bands recorded simultaneously at 2.2--2.3~GHz and 8.0--8.8~GHz. 
Since group delay errors are reciprocal to the frequency range, a wide 
range, 720~MHz, is used in VLBI experiments dedicated to geodesy. 
Radio waves are delayed in the ionosphere, and the magnitude of this delay 
depends on total electron content along the paths --- a highly variable 
quantity. Since the contribution of the ionosphere to path delay is 
reciprocal to the square of effective frequency, simultaneous observations 
at two widely separated frequencies allow the formation of ionosphere-free 
combinations of group delays. 

However for this experiment, the data were recorded in the band 
1368--1432~MHz in dual polarisation (i.e. a single 64 MHz channel). 
Observations within a narrow-band at a low frequency are highly unusual 
for geodetic style experiments since they are not able to provide 
accuracy comparable with dual-frequency, wide-frequency range experiments. 
However, we ran this experiment because in May 2010 the new antennas 
had only L~band  receivers, and this experiment provided an opportunity not
only to make a coarse estimate of station coordinates, but also to test the
equipment. The single frequency setup is a limitation of the VLBI backend 
available at both \askap\ and \wark\ at the time of observations.

\begin{table}
    \caption{\bf The network stations}
    \par\medskip\par
    \begin{tabular}{l @{\qquad} l @{\qquad}  l @{\qquad}  l }
       \hline
       Name            &  \ntab{c}{$\phi_{gd}$\hphantom{aa}}   & 
                          \ntab{c}{$\lambda$\hphantom{aaa}}     & Diam \\
       \hline
       {\askap       } & $-26\Degr.69$  & $116\Degr.64$ & 12 m \\
       {\atca        } & $-30\Degr.31$  & $149\Degr.56$ & 22 m \\
       {\hobart      } & $-42\Degr.81$  & $147\Degr.44$ & 26 m \\
       {\sc mopra    } & $-31\Degr.27$  & $149\Degr.10$ & 22 m \\
       {\sc parkes   } & $-33\Degr.00$  & $148\Degr.26$ & 64 m \\
       {\wark        } & $-36\Degr.43$  & $174\Degr.66$ & 12 m \\
       \hline
    \end{tabular}
    \label{t:lba}
\end{table}

  The experiment was scheduled using {\sf sur\_sked} software program in 
the VLBI geodetic mode. A pool of 113 sources with correlated flux 
densities at X~band  greater than 700~mJy based on results of the ongoing 
LBA Calibrator Survey \citep{r:lcs1} observing campaign was used. 
The scheduling goal was to have a uniform sky coverage at each station. 
This was achieved by putting a next source in the schedule that has 
a maximum score among other candidate sources. The score was assigned 
according to 
\beq
    S = \Frac{1}{t_s} \lp \Frac{D_{min} + 0.1}{0.66} \rp^4 ,
\eeq{e:e1}
  where $t_s$ is {\Note maximum slewing time for all participating antennas} 
in seconds, $D_{min}$ is the distance in radians to any source observed in 
the past. If a source was observed during the previous 150 minutes, it 
was barred from consideration as a candidate. Each scan had a nominal 
duration of 150~s. \Note{Every hour a set of 4 atmosphere calibration scans
was observed: two scans with all antennas with their elevation in the range  
$[12\degr, 45\degr]$, one scan at elevations $[32\degr, 45\degr]$ and one 
scan at elevations $[45\degr, 90\degr]$.} The scheduling algorithm for each 
set found all combinations of calibrator sources that fell in the elevation 
ranges and selected the sequence of four objects that minimized the slewing 
time. The purpose of these calibration observations was 1)~to serve as 
amplitude and bandpass calibrators; 2)~to improve robustness of estimates of 
the path delay in the neutral atmosphere.

\section{Data Analysis}    \label{s:anal}

\subsection{Data transfer and correlation} 

  Following the observations, the data from \atca, \hobart, 
{\sc mopra} and {\sc parkes} were transferred, with the support of 
ARCS (the Australian Research Collaboration Service), via fast network 
connections to a petabyte data store operated by iVEC in Western Australia. 
The data from \wark\ were transferred 
from the telescope using the trans-Tasman link directly to Perth.
The data from \askap\ were transported to Perth by car, as the fast
network to the MRO at the time of the experiment was under construction.
All data were later transferred to disks mounted on the correlation 
cluster at Curtin University and correlated using the DiFX 
software correlator \citep{r:del07}. Initial antenna positions 
for \wark\ and \askap\ determined from GPS receivers were used for 
correlation. Other station positions are known from VLBI solutions with 
accuracy better than one centimeter \citep{r:geo_lba}. The data were 
correlated using an integration period of 0.25 seconds and 512 frequency 
channels across the 64~MHz band. All four combinations of right and 
left circular polarization signals were correlated. 

\subsection{Data quality control made at the correlator}

  Initial data quality checking and clock searching was performed 
at the correlator using an automated pipeline script implemented using 
the ParselTongue AIPS interface \citep{kettenis06}. Hydrogen masers were 
used to provide the frequency standard at all stations except at 
\askap\ where a lower stability (but still acceptable)
Rubidium frequency standard was used. This resulted in a noticeable
correlation loss on baselines to \askap. \Note{All stations}, except 
\wark\ observed with dual circular polarizations, while \wark\
observed dual linear polarizations due to a malfunction of the linear to
circular conversion module, resulting in a reduced SNR  on \wark\ 
baselines: $\sim \sqrt{2}$ less sensitive than expected from a priori
SEFD estimates. {\sc atca} participated as a phased array comprising 5x22-m
dishes. Polarization isolation at {\sc parkes} receiver did not work correctly
during the first 2 hours of the experiment, but this was fixed after the 
linear to circular hybrid module was reset.

\subsection{Post-correlator data analysis}

  The spectrum of cross correlation and autocorrelation for each scan and 
each baseline computed by the correlator at a uniform two-dimensional grid 
of accumulation periods and frequencies was used for further processing. 

  The fringe fitting procedure searches for phase delay $\tau_p$,
phase delay rate $\dot{\tau}_p$, group delay $\tau_g$, and its 
time derivative $\dot{\tau}_g$ that correct their a~priori values used
by the correlator model in such a way that the coherent sum of
weighted complex cross-correlation samples \Note{over a given baseline
and a given scan,} $c_{ij}$
\beq
   \begin{array}{ll}
      C(\tau_p,\tau_g,\dot{\tau}_p,\dot{\tau_g}) = & 
          \dss \sum_i \sum_j c_{ij} \, w_{ij} \times \\ &
          \hspace{-8em}
          e^{i( \omega_0 \tau_p \; + \;
                \omega_0  \dot{\tau}_p (t_i-t_0) \; + \;
                (\omega_j - \omega_0) \tau_g \; + \;
                (\omega_j - \omega_0) \dot{\tau}_g (t_i - t_0) )}
          \hspace{-3em}
     \vex
   \end{array}
\eeq{e:e2}  
  reaches the maximum amplitude. Index $i$ runs over time and index $j$ 
runs over frequencies. $\omega_0$ and $t_0$ denote the angular reference
frequency within the band and the reference time within a scan and $w_{ij}$
are weights that are defined as a fraction of processed samples
in each accumulation period. Software \PIMA was used for amplitude 
calibration and evaluation of phase and group delay, as well as their 
time derivatives. The algorithm implemented in \PIMA is described in 
detail in \citet{r:vgaps}.

\begin{figure}[ht]
   \includegraphics[width=0.48\textwidth]{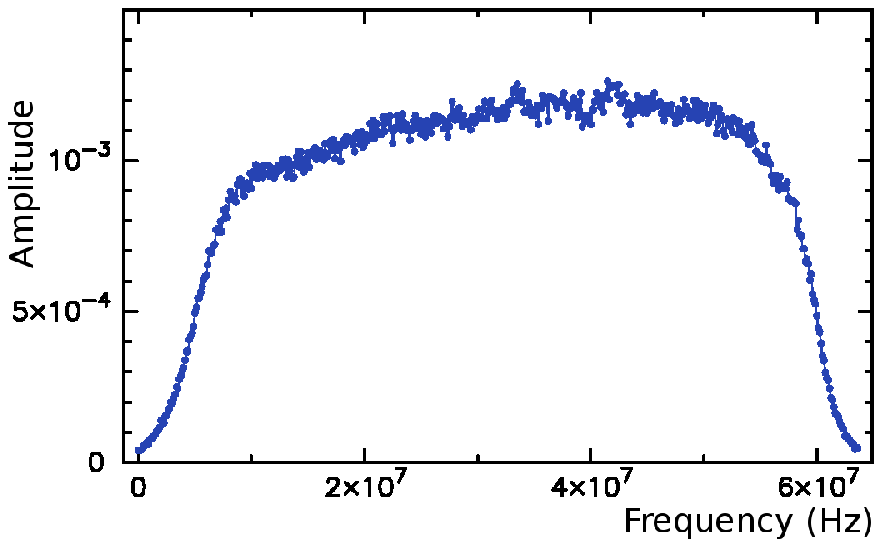} \\
   \par\medskip\par
   \includegraphics[width=0.48\textwidth]{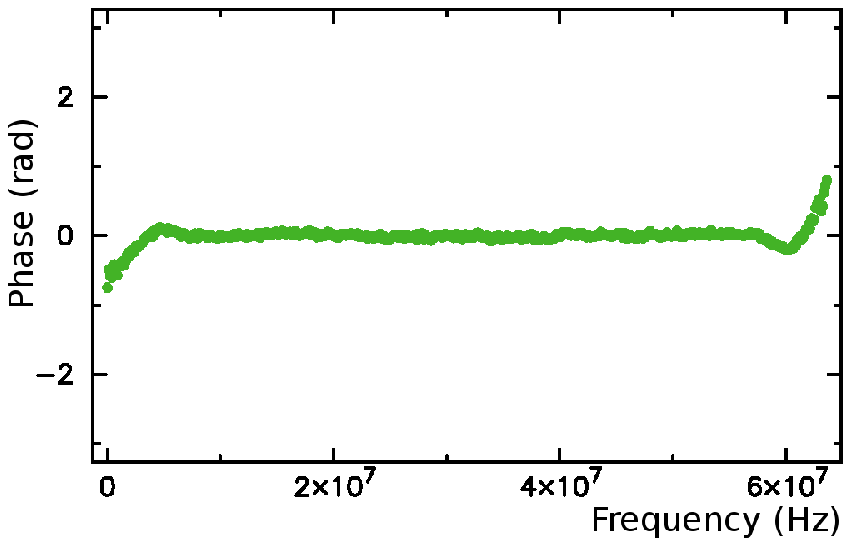}
   \caption{The amplitude (upper plot) and phase (lower plot)
            of the cross correlation spectrum \note{in a scan of
            source 0537$-$441} at baseline 
            \askap/{\sc{parkes}} averaged over time 
            after applying fringe search. \note{Integration time 0.25~s}}
   \label{f:fringe_freq}
\end{figure}

  After computing phase and group delays and their derivatives, the spectrum
of cross-correlations can be averaged over time or frequency. This provides
valuable diagnostics. The amplitude spectrum at new stations \askap\ and
\wark\ shows a significant drop of the amplitude near the edges of 
the band as it is seen in Figure~\ref{f:fringe_freq}. The SNR losses
due to the shape of the filter is at a level of 13\%. A portion of the 
band with the fringe amplitude less than 1/4 of the maximum was filtered
out in the final processing. 

\begin{figure}[ht]
   \includegraphics[width=0.48\textwidth]{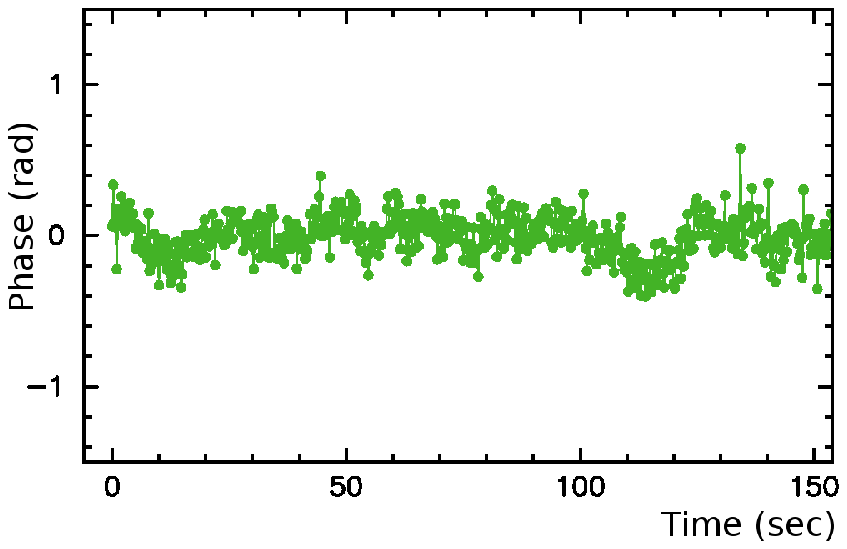}
   \par\medskip\par
   \includegraphics[width=0.48\textwidth]{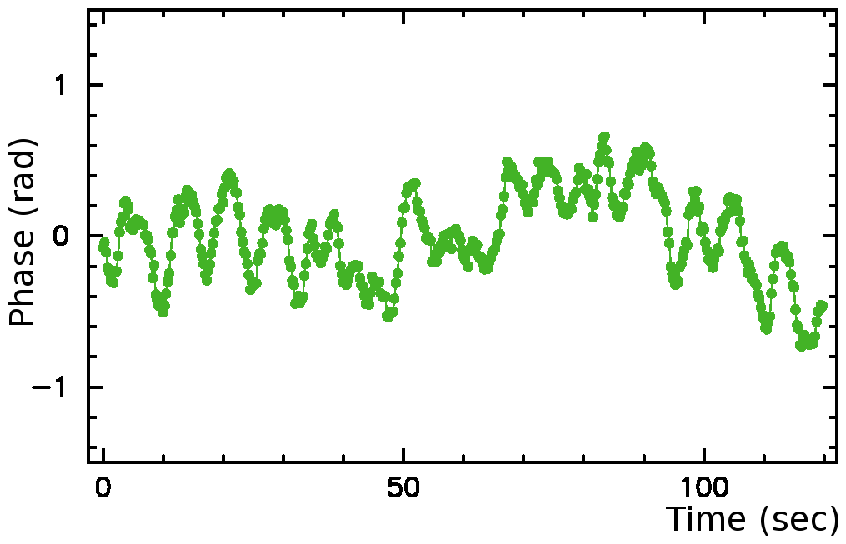} \\
   \caption{The phase of the cross correlation spectrum \note{in a scan 
            of source 0537$-$441} at baseline \wark/{\sc{parkes}} 
            (upper plot) and \askap/{\sc{parkes}} (lower plot) averaged over 
            frequency after applying fringe search solutions.
            }
   \label{f:fringe_time}
\end{figure}

  Since the observations were made during low solar activity, the 
ionosphere did not cause significant decorrelation. Instability
of the Rubidium frequency standard is clearly seen in 
Figure~\ref{f:fringe_time}, \Note{and it caused a de-correlation 
on baselines with \askap\ at a level of 4--5\%}.

  At the first step of geodetic data analysis, the spectrum of the cross-correlation
function was screened. The edges of the bandpass and several spurious peaks 
near the center of the bandpass were filtered out. The complex bandpasses
were evaluated and the fringe-fitting procedure was repeated with the 
the refined bandpasses and with the bandpass mask applied. This procedure 
is described in full detail in \citet{r:vgaps}. Analysis of the amplitudes 
at all combinations of right and left circular polarizations, RR, LL, LR and RL, 
confirmed that the polarization setup at {\sc parkes} for the first two 
hours was incorrect. But it was also found that polarization impurity 
defined as $\sqrt{|RL|^2 + |LR|^2}/\sqrt{|RR|^2 + |LL|^2}$ was 
significant on baselines with other stations. For example, the 
polarization impurity at baseline
\atca/{\sc{mopra}} was at a level 7\%, on baselines \atca/\hobart26,
\atca/\askap\ and \atca/{\sc{parkes}} was 25--30\%,
and the baseline  \atca/\wark\ was $\sim\! 100\%$.
Therefore, the coherent combining dual-polarization improved the SNR
at some baselines and degraded at others. For this reason, the datasets
of group delays with RR and LL polarizations were used as independent
experiments in further processing.

\begin{table*}[t]
   \caption{Coordinates of \askap\ and \wark\ derived from
            analysis of experiment vt14a. \hspace{0.2\textwidth} }
   \label{t:coord}
   \par\vhx\par
   \begin{tabular}{l @{$\quad$} r @{$\quad$} r @{$\quad$} r}
       \hline
       Station       & \nnntab{c}{Station coordinates (m)} \\
                     & \ntab{c}{X} & 
                       \ntab{c}{Y} & 
                       \ntab{c}{Z} \vex \\
      \askap & $ -2556741.89  \pm  0.09 $ & 
                       $  5097441.23  \pm  0.14 $ &
                       $ -2847748.34  \pm  0.11 $ 
                                      \hphantom{aaa} \vhx \\
      \wark & $ -5115325.55  \pm  0.10 $ &
                       $   477842.95  \pm  0.05 $ &
                       $ -3767194.41  \pm  0.09 $ 
                                      \hphantom{aaa} \vhx \\
       \hline
   \end{tabular}
\end{table*}

  The next step of the analysis pipeline is to compute theoretical path 
delays and form small differences between them and derived group delays.
Computation of theoretical path delays was made with the use of the
{\sf VTD} software program\footnote{{\tt http://astrogeo.org/vtd}} that 
incorporates the state-of-the art geophysical models.
Our computation of theoretical time delays in general follows the approach 
of \citet{r:masterfit} with some refinements. The most significant 
ones are the following. The advanced expression for time delay 
derived by \citet{r:Kop99} in the framework of general relativity 
was used. The displacements caused by the Earth's tides were computed 
using the numerical values of the generalized Love numbers presented 
by \citet{r:mat01} following a rigorous algorithm described by
\citet{r:harpos} with a truncation at a level of 0.05~mm. 
The displacements caused by ocean loading were computed by convolving 
the Greens' functions with ocean tide models. The GOT99.2 model of diurnal 
and semi-diurnal ocean tides~\citep{r:got99}, the NAO99 model \citep{r:nao99} 
of ocean zonal tides, the equilibrium model \citep{r:harpos} of the pole tide, 
and the tide with period of 18.6 years were used. Station displacements 
caused by the atmospheric pressure loading were computed by convolving 
the Greens' functions that describe the elastic properties of the 
Earth \citep{r:farrell} with the output of the  atmosphere NCEP Reanalysis 
numerical model \citep{r:ncep}. The algorithm of computations is described 
in full detail in \citet{r:aplo}. The displacements due to loading caused 
by variations of soil moisture and snow cover in accordance with GLDAS Noah 
model \citep{r:gldas} with a resolution $0.25^\circ \times 0.25^\circ$ 
were computed using the same technique as the atmospheric pressure loading. 
The empirical model of harmonic variations in the Earth orientation parameters 
{\tt heo\_20101111} derived from VLBI observations according to the method 
proposed by \citet{r:erm} was used. The time series of UT1 and polar motion 
derived by the NASA Goddard Space Flight Center operational VLBI solutions 
were used a~priori. 

  The a~priori path delays in the neutral atmosphere in the direction of 
observed sources were computed by numerical integration of differential 
equations of wave propagation through the heterogeneous media. 
The four-dimensional field of the refractivity index distribution was 
computed using the atmospheric pressure, air temperature and specific 
humidity taken from the output of the Modern Era Retrospective-Analysis for 
Research and Applications (MERRA) \citep{r:merra}. That model presents 
the atmospheric parameters at a grid 
$1/2\degr \times 2/3\degr \times 6^h$ at 72 pressure levels.

  In the initial least square (LSQ) solution, positions of all stations, 
except {\sc parkes}, were estimated, as well as coefficients of the expansion 
of clock function and residual atmosphere path delay in zenith direction 
into the B-spline basis of the 1st degree. The quality check revealed 
a clock break at station \hobart. During the preliminary phase of the 
data analysis, outliers were eliminated and the baseline-dependent 
corrections to the a~priori weights defined to be reciprocal to 
formal uncertainties of group delays were determined in such a way that
the ratio of the weighted sum of squares of residuals to their mathematical
expectation was close to unity. Group delays from RR and LL polarization
data were processed independently.

  The final LSQ solution used all VLBI group delays collected from 1984 
through \Note{2010\footnote{Avaible at the IVS Data Center at \hfill \\
{\tt http://ivscc.gsfc.nasa.gov/products-data/index.html}}}, in total, 
7.5 million values, including \Note{824 group delays}
from this experiment. Positions of all stations, all sources, the 
Earth orientation parameters as well as over 1 million  nuisance parameters 
were estimated in a single LSQ run. Minimal constraints were imposed 
to require that net-translation and net-rotation over new position
estimates of 48 stations with long history with respect to positions 
of observations to these stations in the ITRF2000 catalogue 
\citep{r:itrf2000} to be zero. This ensures that positions of all 
stations, including \askap\ and \wark\ be consistent 
with the ITRF2000 catalogue. More details about parameter estimation 
technique can be found in \citet{r:rdv}. 

  The estimates of coordinates of \askap\ and \wark\
on epoch 2010.05.07 are given in table~\ref{t:coord}. The errors 
reported in the table are the formal uncertainties from the LSQ
solution computed in accordance with the error propagation law.
Since there was only one experiment available with rather an unusual setup, 
it is difficult to provide a realistic estimate of errors. 

  The largest source of systematic error is the path delay in the 
ionosphere. Our attempt to use maps of the total electron content
from GPS using the data product from the analysis center CODE
did not improve the fit and shifted estimates of station coordinates 
at a fraction of the formal uncertainty \Note{($0.2\sigma$)}. Analysis of 
dual-band VLBI
experiments showed that global TEC maps above Australia and 
New Zealand provided by the International Global Navigation Satellite 
System Service are not reliable, and the path delay computed 
from this model poorly represents the true delay in 
the ionosphere, at least during solar minimum \citep{r:lcs1}.

  The adjustments to the residual zenith path delay in the atmosphere 
were in the range of 1--5~ns, while they are typically in the range 
0.03--0.1~ns for dual-band observations. Unlike the estimation of the 
mismodeled troposphere path delay, estimation of unmodeled ionosphere 
path delay in zenith direction from the observations themselves does 
not adequately represent the true path delay, since when antennas point 
at different directions, the ionosphere piercing points with a typical 
height $\sim\! 450$~km may be located at distances up to 1000~km.

  For evaluation of the robustness of our coordinate estimates we made 
a trial solution and estimated the position of station \atca\
independently. Estimates of the of \atca\ position appeared 
within $2\sigma$ of the CATN5 pad. The position of CATN5 was derived 
from the position of the pad CATW104 determined from previous VLBI 
observations and from the results of a local survey that measured 
coordinates of CATN5 with respect of CATW104 \citep{r:geo_lba}. This 
gave us a hint that the reported uncertainty may be underestimated by 
a factor of~2, especially for the vertical component. In our final 
solution we applied the eccentricity vector between CATN5 and CATW104 
from the local survey.

  The data from RR and LL polarizations have independent random errors, but
they share the same systematic errors. In another trial solution we processed
RR and LL data separately. The position differences of these two solutions
were within $1\sigma$ for horizontal component and $2\sigma$, or 0.4~m
for the vertical component.

\begin{figure}[ht]
   \includegraphics[width=0.48\textwidth]{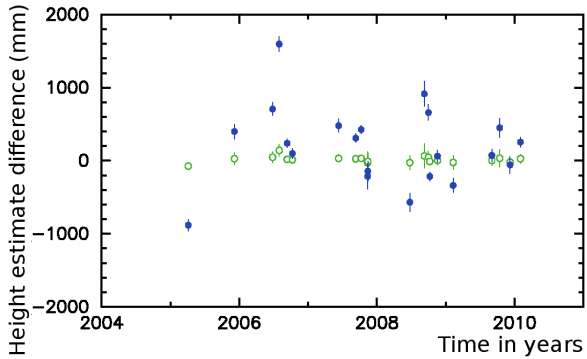}
   \caption{The differences in the vertical coordinate estimate
            of station \hobart\ from a series of trial
            solutions, when position was estimated only from
            one experiment at a single baseline with {\sc parkes}.
            Hollow circles represent the difference between
            the X~band  only solution and the X/S dual band
            solution. The \note{weighted root mean square (wrms)} 
            is 40~mm. Filled discs represent 
            the difference between the S~band  only solution and 
            the X/S dual band solution. The wrms is 490~mm. }
   \label{f:height_diff}
\end{figure}

  However, the latter test is not sensitive to common systematic errors 
caused by unmodeled ionosphere path delays. In order to evaluate
the magnitude of the ionosphere path delay errors, we used prior 
observations at the baseline {\sc parkes}/\hobart\ under the IVS 
geodetic program. We selected 21 twenty four hour experiments in 2005--2010
--- the period of solar minimum. We ran three sets of 21 trial solutions.
Within each set we used all experiments from 1980 through 2010 with station
\hobart\ removed from all but one experiment. The only experiment 
with \hobart\ was different in each solution within a set. In such 
an experiment all other stations, except \hobart\ and {\sc parkes} were 
removed. This setup emulated the case of determining station coordinates from 
one experiment only. The reference set $A$ used an ionosphere-free 
combination of X and S band group delays. The set $B$ used X~band  group delays
for an experiment with \hobart\ and X/S combinations for all other
experiments. The set $C$ used S~band  group delays. We got three series of 21
estimates of \hobart\ positions. We formed the difference between
estimates of \hobart\ vertical and horizontal coordinates between
sets $B$ and $A$, and between sets $C$ and $A$. The differences for vertical 
component of site position estimates are given in Figure~\ref{f:height_diff}.

  The only difference between solutions in sets $A$, $B$, and $C$ is the 
treatment of the ionosphere path delay. Therefore, we can consider
the difference $B-A$ and $C-A$ are primarily due to the contribution 
of unmodeled ionosphere path delay into estimates of site position. 
The average ratio of differences $(B-A)/(C-A)$ is 13.7 which is very 
close to the square of the ratio of effective ionosphere frequencies at 
X and S bands: $(8.387/2.250)^2 \approx 13.9$. This supports our 
argument that the differences $C-A$ are due to the ionosphere path 
delay at S~band. Since the differences $B-A$ and $C-A$ are nicely scaled 
with the square of effective frequency, we can extrapolate the estimates of 
the \Note{weighted root mean square (wrms)} of site position time series 
from solution $C$ to L~band by scaling them by $(2.250/1.4)^2 \approx 2.6$. 
If we were determining the position of \hobart\ from one of these 
twenty-four experiment at L~band, the wrms of position errors due to 
unaccounted ionosphere path delay would have been 1.3~m for the vertical 
component and 0.14~m for the horizontal component. We extrapolate these 
numbers to our estimates of \askap\ and \wark\ positions.

  The differences in position estimates of \wark\ from VLBI and 
from GPS surveys, VLBI--GPS, are 1.66~m for the vertical component, 
0.49~m for the east component, and $-0.48$~m for the north component.
The difference in the vertical component can be explained by the effect
of the ionosphere.

\section{Summary and Future Work}      \label{s:concl}

  We obtained the first estimates of the positions of \askap\ and 
\wark\ antenna reference points from VLBI observations. 
The random position $1\sigma$ errors, 5--6~cm for the horizontal 
coordinates and 20--40~cm for the vertical component, are close to that 
what one can expect from narrow-band observations at L~band. It is 
a pleasant surprise that the very first observations that followed 
the successful fringe test yielded a reasonable result. However, the 
systematic errors due to unaccounted ionosphere are significantly 
greater: 1.3~m for the vertical component and 0.14~m for the horizontal 
component. We also identified several problems with station equipment 
that will be fixed in the future. \Note{The use of a Rubidium frequency
standard at \askap\ caused a decorrelation at a level of 4--5\% within 
2 minute long scans. The frequency instability at longer time intervals
appeared negligible with respect to errors caused by the ionosphere.}

  The existing L~band (1.4~GHz) receiver on \askap\ will be replaced 
\Note{in early 2011} with a new receiver comprising a purpose-built feed 
horn optimized for this antenna, with significantly higher efficiency 
and a noise calibration source. In 2011 \askap\ may be equipped with 
an X~band (8.4~GHz) receiver, and the possibility of an upgrade 
to a higher frequency receiver is being considered.  

  \Note{In 2011, \wark\ was equipped with S/X dual-band receivers. Future
geodetic experiments at these frequencies are planned for February 2011.
The participation of stations \askap\ and \wark\ in VLBI experiments
under absolute astrometry and geodesy programs will permit an
improvement in the precision of results by two orders of magnitude and
reach a millimeter level of accuracy. A more precise survey to provide
1--3 mm accuracy of the tie vector between VLBI and GPS antenna
reference points is planned. Combined with further VLBI observations it 
will then be possible to reconcile the currently observed differences
between GPS and VLBI positions of the \wark\  antenna reference point.}

  In November 2010, Telecom New Zealand made its 30-metre satellite 
Earth station available to AUT's Institute for Radio Astronomy and 
Space Research. It is a wheel-and-track beam-waveguide antenna built 
by NEC Corporation in 1984.  It will be converted to a radio telescope 
capable of conducting both astronomical and geodetic research. 

\section*{Acknowledgments} 

This work uses data obtained from the Murchison Radioastronomy Observatory 
(MRO), jointly funded by the Commonwealth Government of Australia and 
the Western Australian State government. The MRO is managed by the CSIRO. 
We acknowledge the Wajarri Yamatji people as the traditional owners 
of the Observatory site.  
The Long Baseline Array is part of the Australia Telescope National Facility 
which is funded by the Commonwealth of Australia for operation as 
a National Facility managed by CSIRO.
The International Centre for Radio Astronomy Research is a Joint Venture 
of Curtin University and The University of Western Australia, funded by 
the Western Australian State government.
iVEC is a joint venture between CSIRO, Curtin University, Edith Cowan 
University, The University of Western Australia and Murdoch University 
and is supported by the Western Australian Government. 
SJT is a Western Australian Premier's Research Fellow, funded by 
the Western Australian government. 
This project was supported in part by the Australian Research Collaboration 
Service (ARCS). 
We used in our work the dataset MAI6NPANA provided by 
the NASA/Global Modeling and Assimilation Office (GMAO) in the framework 
of the MERRA atmospheric reanalysis project.

\end{document}